\documentstyle[pra,aps]{revtex}
\begin{document}
\draft
\title{The Einstein-Hopf model within the realm of stochastic electrodynamics}

\maketitle
\begin{center}
 O. A. Senatchin
\end{center}

\begin{center}
 Institute of Theoretical and Applied Mechanics
\end{center}

\begin{center}
 SB Academy of Sciences, Institutskaia St. 4/1,
\end{center}
\begin{center}
 Novosibirsk, Russia 630090
\end{center}

\begin{center}
 E-mail: olsenat@itam.nsc.ru ~ and ~ olsenat@mail.ru
\end{center}

\begin{abstract}

The fundamental Einstein-Hopf work that convinced the most part of
physicists, since it had appeared, to take quantum ideas seriously, is
reanalysed in this paper. We have studied the genesis of the work and have
found the conclusion made by Einstein and Hopf not so unambiguous. There
may exist classical, not quantum interpretations of the result. T. H. Boyer
proposed one of those in 1969. Unfortunately, his derivation contains
a loophole in its argument. We propose here a new classical derivation of
Planck's radiation law, with the Einstein-Hopf model being used.

\end{abstract}
\indent

\section{Introduction}

According to Einstein's own statement [1], quantum problems were central
to his activity. The photoeffect, the specific heat of solids,
the spontaneous and induced emission, the Bose-Einstein statistics of gases
--- his most known results --- all were obtained by making use of quantum
ideas. More surprising, as it seems to many, was his following rejection of
traditional quantum physics in a form of wave or matrix mechanics in the midst
of 1920's and later. However, there are few, if any who took note of the fact
that almost in a half of early Einstein's works he tried to derive quantum
effects (in particular Planck's radiation law) from the standpoint of
classical physics and he achieved a great deal.

In his primary analysis of Planck's works Einstein [2-4] attracted an
attention to the contradiction between the ``electrodinamic'' part and
the ``statistical" part of Planck's considerations. Namely the derivation of
equation

\begin{equation}
\rho(\omega,T)=\frac{\omega^2}{\pi^2 c^3}E(\omega,T)
\end{equation}
from Maxwell's electrodynamics was inconsistent with the derivation of
the oscillator energy expression

\begin{equation}
E(\omega,T)=\frac{\hbar\omega}{e^{\hbar\omega/kT} -1},
\end{equation}
which is based not upon a classical mechanics and does not have even a correct
classical limit, i.e., $E(\omega,T)\neq kT/2$ at $kT\gg\hbar\omega$.

If the classical mechanics is not true, writes Einstein, why the Eq. (1),
which was obtained by making use of equation

\begin{equation}
m\frac{d^2 x}{d t^2}+m\omega_0^2 x -\Gamma\frac{d^3 x}{dt^3}=eE(x,t),
\end{equation}
must be correct. The Eq. (3) is based upon classical mechanics and classical
electrodynamics. Herein $\omega_0$ --- is natural frequency of an oscillator,
$\Gamma$ --- is a radiation damping constant, $E(x,t)$ --- is a stochastic
electric field having a Gaussian distribution and named by Planck
``natural" radiation.

Essentially, Einstein pointed out that the expression for the number of
modes in a unit volume --- $\omega^2/\pi^2 c^3$, (the correct
derivation of which is the Eq. (1)) may be erroneous. This remark will be
very important for our further considerations, but
now we are going to reconstruct Einstein's train of thought. So, he put in
question the validity of
classical electrodynamics, the correctness of Eq. (3), and suggested that
processes of absorption and emission of radiation should be analysed on a
different, more general basis. For early Einstein it implied addressing to
purely thermodynamic reasoning and looking for an energy balance equation,
by analogy with the fluctuation-dissipation relation found by him earlier
for the Brownian motion. In this case an emission rate could be specified
by radiation friction of an oscillator, and an absorption rate --- by the
expression for mean-square fluctuations of radiation. Earlier, in the work [5],
he found from thermodynamics the following general formula for the latter:

\begin{equation}
\langle\delta E^2\rangle=kT^2\frac{d\rho(\omega,T)}{dT}.
\end{equation}
In his work [6], Einstein substituted the Planck radiation law

\begin{equation}
\rho(\omega,T)=\frac{\omega^2}{\pi^2 c^3}\frac{\hbar \omega}
{e^{\hbar\omega/kT} -1},
\end{equation}
according perfectly with experimental data, in (4) and yielded

\begin{equation}
\langle\delta E^2\rangle=\hbar\omega\rho +\frac{\pi^2 c^3}{\omega^2}\rho^2.
\end{equation}
If it was possible to understand how to obtain from electrodynamics the
similar expression, containing two terms, then it would be possible to
reverse the consideration by substituting (6) into fluctuation-dissipation
relation

\begin{equation}
\langle \delta E^2\rangle\simeq R kT,
\end{equation}
and after determing, also from electrodynamics, the resistance factor $R$, to
find the Planck radiation law in a framework of classical physics. Indeed,
by using  special relativity formulas, Einstein succeeded in calculating $R$
for simple model of a moving mirror. He also determined that it was
possible to obtain the second term of (6) ``on dimentional grounds" from
Maxwell's electrodynamics as a consequence named the ``wavelike" term. But the
first term gave rise to a real problem.

Einstein, just as the great part of physicists of his time, was
fascinated by Boltzmann's kinetic theory of gases, and it seemed to him that
the most natural interpretation of the term was to imagine the radiation as
consisting of isolated pointlike objects --- light quanta, or photons.
Therefore, he named the term the ``particlelike" term.
Although Einstein himself was the author of this interpretation, he was not
absolutely sure about it. In fact, if we put not the Planck radiation law,
but the Rayleigh-Jeans law in (4):

\begin{equation}
\rho(\omega,T)=\frac{\omega^2}{\pi^2 c^3}\frac{kT}{2},
\end{equation}
then we will obtain the second, ``wavelike" term of Eq. (6). Analogously, if
we put the Wien law in (4):

\begin{equation}
\rho(\omega,T)=\frac{\omega^2}{\pi^2 c^3}e^{-\hbar\omega/kT},
\end{equation}
then we will obtain the first, ``particlelike" term of Eq. (6). Einstein could
hardly avoid noting of it, and his constant attention to the Wien radiation
law is the evidence. Suffice it to mention his works [2] and [9].

However, as it is well known, Wien's law was derived by taking into
account the Doppler shift of frequency at reflection of radiation from a moving
piston. This is quite possible that for finding the first term of Eq. (6)
within
a framework of classical electrodynamics it would be sufficient to take into
account carefully a uniform motion of a mirror immersed into radiation. That
is why Einstein wanted to do the calculations more thoroughly in order to come
to final conclusion.

At Salzburg conference of 1909, where Einstein reported the ideas [3], he
met Ludwig Hopf, a brilliant mathematical physicist, who made his doctoral
degree not long before under the direction of Professor Sommerfeld [7].
As a result of their joint efforts, the famous Einstein-Hopf work [8] appeared.
We are going to discuss it now more thoroughly than in the usual
practice [10-13].

\section{The Einstein-Hopf model}

In their article Einstein and Hopf returned once again to Planck's
oscillator, but to make it movable, just like Einstein's mirror [3,6], they put
the oscillator inside of a gas molecule. After works of Planck, not many dared
to apply the energy equipartition theorem to radiation or to particle inner
degrees of freedom. However, as Einstein and Hopf writes, ``the applicability
of the theorem [to a molecular translational motion] is proved sufficientely
by kinetic theory of gases to leave no doubt". Hence, the authors decided to
go along the way, opposite of Planck's, and invoked the equipartition law.
To do so, they wrote down one version of a fluctuation-dissipation
relation justified for a particle immersed in a thermal electromagnetic
radiation:

\begin{equation}
\langle\Delta^2\rangle=2PkT\tau,
\end{equation}
where $\langle\Delta^2\rangle$ --- is a mean-square impulse, $P$ --- a
friction coefficient.The authors calculated those values by using classical
electrodynamics and, therefore, by making use of Planck's Eq. (3).
The substantial
difference is that since an oscillator moves, therefore, the motion equation
is no more one-dimensional. Besides, the motion of a particle relative to an
electric field brings into existence a magnetic field, hence, we must take
into account both fields. So, if we assume that the dipole oscillator is
oriented along the z-axis and the particle is constrained to move along the
x-axis, we can write down the following set of equations:

\begin{eqnarray}
m\frac{\partial^2 z}{\partial t^2} +\omega_0^2 z -\Gamma\frac{\partial^3 z}
{\partial t^3} =e\left(E_z +\frac{1}{c}B_y\frac{\partial x}{\partial t}\right)
\enskip
\nonumber\\
m\frac{\partial^2 x}{\partial t^2} +\omega_1^2 x -\Gamma\frac{\partial^3 x}
{\partial t^3} =e\left(E_x +\frac{1}{c}B_y\frac{\partial z}{\partial t}\right).
\end{eqnarray}
Since the velocity of a particle along the x-axis is small compared to the
speed of light $c$, we can omit the second term in the right hand side of the
first equation. By definition the oscillator vibrates only along the z-axis,
therefore $\omega_1$=0. Furthermore, it is in a thermodynamical equilibrium
with
radiation, therefore, its acceleration is small, and we can neglect
radiation damping force along the x-axis. Then the set of equations (11)
goes to

\begin{eqnarray}
m\frac{\partial^2 z}{\partial t^2} +\omega_0^2 z -\Gamma\frac{\partial^3 z}
{\partial t^3} =eE_z \enskip\quad\,
\nonumber\\
m\frac{\partial^2 x}{\partial t^2}=e\left(\frac{\partial E_x}{\partial z}z +
\frac{1}{c}B_y\frac{\partial z}{\partial t}\right).
\end{eqnarray}
This is the set of equations that were solved by Einstein and Hopf. First,
they found $z$ from the first equation and then substituted it in the second
equation and reduced it. The friction force appeared to be proportional, as
in the Brownian particle problem, to the velocity of a particle:

\begin{equation}
\langle F\rangle=m\frac{\partial^2 x}{\partial t^2}=Pv,
\end{equation}
where the friction coefficient $P$ is equal to
\begin{equation}
P=c \pi^2\frac{6}{5}\Gamma\left[\rho(\omega_0)-\frac{\omega_0}{3}\frac{d
\rho(\omega_0)}{d\omega_0}\rho(\omega_0)\right].
\end{equation}
The mean-square impulse is calculated by squaring the impulse $\Delta=m\dot x$
and averaging by using the properties of stochastic radiation field. The
result is

\begin{equation}
\langle\Delta^2\rangle=\frac{4\pi^4  c^4 \Gamma}{5\omega_0^2}\rho^2(\omega_0).
\end{equation}
To be precise, we must point out that Einstein and Hopf considered only a
resting oscillator for simplification of calculations. However, we are
inclining to think that they did the calculation of $\langle\Delta
^2\rangle$ for a moving oscillator, too. It was typical for manner of Einstein
--- to calculate all possible, but to present in articles the simplest
necessary derivation, not burdening a consideration with details that give
nothing new in principle.

However, all efforts were in vain. They failed to catch the
``particlelike" term by this net. More recently Boyer [14] suspected that the
approximations were not sufficient to solve the problem, and that
full relativistic transformations of all important values may be necessary.
In work [15] Boyer did the entire calculations, but obtained, however,
the result anologous to that of Einstein and Hopf. In the case the
``particlelike" term escaped also. There is nothing for it but to wonder,
once again, of the power of intuition that Einstein had.

\section{Zero-point energy and a concise history of stochastic electrodynamics}

Thus, in 1910, the idea of photons became preferable, and since then
the Einstein-Hopf work has been considered as a turning point in formation of
the new quantum theory. However, two years passed, Planck introduced
the ``zero-point" energy into physics [16], and in a new Einstein's work (with
Otto Stern) [17] the ``particlelike" term was interpreted in a different
fashion as a term connected with the zero-point energy. Thereby, at last,
the Planck radiation law was derived within a framework of classical physics.
Afterwards, Einstein abandoned such interpretation of the radiation law
[18], specifically, because of lack of a reasonable explanation for a
factor 2, arising therewith.

In his profound scientific-historical research [19] Milonni shows
that a resolution of this factor of 2 discrepancy is not taking into account
by Einstein and Stern the zero-point energy of electromagnetic radiation, in
spite of their considering the zero-point energy of the oscillator. If
the calculations are done correctly, in the mean-square impulse one more,
additional term arises:

\begin{equation}
\langle\Delta^2\rangle=\hbar\omega\rho +\frac{\pi^2 c^3}{\omega^2}\rho^2 +
\frac{\hbar\omega}{2}\rho_0(\omega).
\end{equation}
It was clear from the very beginning to Timothy Boyer in his first work devoted
to the blackbody radiation law [20]. He ascribed this term to the effect of
a particle interaction with cavity walls. If some part of an oscillator energy,
regardless of temperature, is lost all the time, due to transferring it to
walls, it could
be possible to account the balance of energy between the oscillator and
the radiation
correctly. By combining thermal and zero-point fields $\rho=\rho_T +\rho_0$,
Boyer was able to write down the fluctuation-dissipation relation as

\begin{equation}
\frac{1}{3}\frac{c^3 \pi^2}{kT\omega^2}\left[\rho^2(\omega,T) -\left(\frac{
\hbar\omega^3}{2\pi^2 c^3}\right)^2\right]=\rho(\omega,T)-\frac{1}{3}\omega
\frac{d}{d\omega}\rho(\omega,T),
\end{equation}
which has Planck's radiation law with the additional zero-point term as its
solution:

\begin{equation}
\rho(\omega,T)=\frac{\omega^2}{\pi^2 c^3}\left(\frac{\hbar\omega}{e^{\hbar
\omega/kT} -1} +\frac{\hbar\omega}{2}\right).
\end{equation}
Thus, the old problem was presented in a new fashion.

But even if we do not give due attention to the Boyer work, it is necessary
to mention that in 60's it was a revival, Second Coming of Planck's idea about
the zero-point energy. Beginning with fundamental works of T. W. Marshall
[21-23] the idea of zero-point radiation was used to explain effects of
diamagnetism, paramagnetism, Van der Waals forces. The Casimir force [24,24],
the Lamb shift [26-28] were explained, the connection
between a behavior of classical and quantum oscillator was found [21,29].
It gave, at last, an insight into the stability of matter [30,31].
The attempts were made to study Comptom effect [32], spin [33] and hydrogen
atom [34] from the standpoint of classical physics. It is easy to feel
the euforia of protogonists of zero-point radiation field, so called
``stochastic electrodynamics" (SED), when reading Boyer's review
[30] of 1975. It is also important to say that these events were
going against a background of appearance and development of conjugate
with SED
Nelson's stochastic mechanics [35-37] in which the Schrodinger equation
was derived, as it seemed then, from classical mechanics.

However, after some time, the excitment went down. In thermodynamic
balance equation the Maxwell-Boltzmann distribution for particles led
inevitably to the Rayleigh-Jeans distribution for a field. In ``classical"
derivations of Planck's radiation law [20,38-40] irreparable
loopholes were found [41,42]. Specifically, in the mentioned before first
Boyer's derivation
[20], the blackbody radiation should inevitably have a spatial inhomogenuity,
i.e., wall effects. But they were not observed in experiments [43]. Moreover,
at the absence of surrounding walls, the equations led, as it turned out,
to inevitable
conclusion: the zero-point radiation will eventually bring the particles to
high velocities, ever closer to the speed of light! It is most obvious for the
case of absolute zero temperature. Indeed, in the case a particle absorbs
energy from zero-point radiation all the time,
and a friction is absent completely. Only a few could accept the conclusion
as a really existing and tried to find experimental evidences for it,
in particular, in velocity
distribution of primary cosmic rays [44,45]. Others defended persistently the
importance of interaction with walls [46]. For us, however, it is not clear
why a free particle radiates at the moment of contact with a wall, while wall
molecules at the time only absorb. If we have, for example, two elastically
colliding electrons, then first accelerates in one direction and second in
opposite direction. It seems obvious that the electromagnetic fields of these
electrons should be combined together and compensate each other.

\section{New solution of the problem}

Such indefinite status SED had for rather long time. On the one hand, its
methods have found applications in new areas of researches such as gravitation
theory [47], inertial mass theory [48-50] or in an explanation of a nature of
the de Broglie wave [51-53]. On the other hand, it seemed impossible to
find a connection between Maxwell's distribution for particles and Planck's
distribution for a field. However, as it seems to the author of the present
paper, after the historic-critical investigation above, there is an
obvious
opportunity of one more attempt to get rid of the internal contradictions
in SED now. This possibility was founded as early as in
the Einstein-Hopf work. The point is that the Eqs. (11) describes not an
interaction of radiation with an oscillator as a whole, but an interaction
of radiation with a charge $e$! And the charge inside of the oscillator
\em is accelerated \em relative to the electromagnetic field. Moreover, for
example, the acceleration of electron in classical hydrogen atom is very great,
about $10^{25} cm/c^2$. It is difficult to believe that such huge accelerations
do not give any sensible physical effects. Therefore, it would be more
correct to solve the set of equations (12) by taking into account \em an
accelerating motion \em relative to electric and magnetic stochastic fields,
also, not only a uniform motion or a rest state, as it was made by Einstein and
Hopf. Let us notice, in addition, that a charge in a harmonic potential is
uniformly accelerated with the acceleration $a=\omega_0^2 r_0$. Hence, for the
further study of the problem we can use the same mathematical technique that
was made for explanation of Unruh-Devies effect in a framework of SED [54,55],
i.e. the appearance of additional thermal radiation around the observer
accelerating in zero-point radiation. Moreover, in work [48] the necessary
for us calculations of average force $\langle F\rangle$ have already been
done, however, for the case of presence zero-point radiation only. Therefore,
it is possible for us not to
overload the paper with long calculations (and their complexity may understand
everyone who is familiar with the work [48]), but we may take advantage of
its results to do a simple estimation, showing that the occurence of the
second term in the left hand side of Eq. (17) is not only possible, but
quite natural.

So, the authors of the article [48] generalized the Einstein-Hopf model
for the case of uniformly accelerating particle motion and neglected a
thermal radiation. As a result, instead of Eq. (13) they yielded

\begin{equation}
\langle F\rangle=\frac{\hbar\omega_0^2}{2\pi c^2}\Gamma a=m_i a.
\end{equation}
The quantity $m_i$ was interpreted by Haisch, Rueda and Puthoff as a real,
observable inertial mass in distintion to $m$ of Eqs. (12), which is not
observable in experiments.

If we assume that the equation (19) is correct, taking into account the impact
of thermal radiation, we will obtain for average force the following
expression:

\begin{equation}
\langle F\rangle=Pv+m_i a=c\pi^2\frac{6}{5}\Gamma\left[\rho-\frac{1}{3}
\omega_0\frac{d\rho}{d\omega_0}\right]v+\pi c\Gamma\left(\frac{\rho_0}{\omega
_0}\right)\cdot a,
\end{equation}
Where $\rho_0=\hbar\omega^3/2\pi^2 c^3$ is the density of the zero-point
radiation.

Analogously, if we take a uniformly accelerated motion into account,
the mean-square
impulse will have one more, additional term, and it will be proportional
to quantity

\begin{equation}
\langle\Delta^2\rangle_a\sim\Gamma\left(\frac{\rho_0}{\omega_0}\right)^2\sim
\Gamma\hbar^2\omega_0^4.
\end{equation}
The second term of Boyer's equation (17) has exactly such a dependence on
$\Gamma,\hbar$ and $\omega_0$. As for the minus sign before the term, it
follows from physical considerations. An accelerating particle has an
additional thermal field around it (from its viewpoint), i.e., the resistance
to the motion increases. But this effect may be imaginary, existing only
for the particle. From the point of view of an external observer, the particle
may lack the energy from the radiation. From energetic balance reasoning
it is equivalent. And since in Eq. (10) only $P$ is taken into account, and
$m_i$ is absent, we may consider that the basic difficulties are removed.

\section{Conclusion}

History of modern physics demonstrates that any great work is dialectical
by its essence. If it affected physicists once, early or late denying
its conclusions will again inflame the scientific community. The Einstein-Hopf
work in a time of its appearance clearly demonstrated the necessity of
abandoning classical physics and the importance of quantum ideas.
Concepts of
photons and wave-particle duality became a part of annals of physics.
Nevertheless, this was the Einstein-Hopf work that included the mathematical
technique which became a foundation of stochastic electrodynamics --- the
classical theory that explains all quantum phenomena by making use of idea of
classical zero-point radiation. In 1969 T. H. Boyer demonstrated how to derive
Planck's radiation law within a framework of classical physics. Thus, the
problem that Einstein and Hopf were not able to manage was solved.
Unfortunately, one Boyer's assumption about radiation of particles at the time
of their hitting the walls eventually turned out to be inconsistent.
Therefore, the vital question of SED was to offer an alternative
mechanism of energy balance between particles and blackbody radiation. In the
present work such mechanism is proposed. It is an account of accelerating
(relative to radiation) motion of charges inside particles. In the case
an additional term appears in the fluctuation-dissipation relation, which
coincides with the term in Boyer's equation, attributed earlier to the cavity
walls effect. Thus, one the basic contradictions of SED is resolved and a new
way for its further development is found.

\end{document}